# The Energy Navigator – A Web-Platform for Performance Design and Management

*Plesser, Stefan[1], Claas Pinkernell[1], M. Norbert Fisch[2], Bernhard Rumpe[3]*

[1] synavision GmbH, Aachen
Schönauer Friede 80
D-52072 Aachen
Tel. +49 (241) 80 - 21301
Fax +49 (241) 80 - 22218
Germany
{plesser, pinkernell}@synavision.de
http://www.synavision.de

[2] Institute of Building Services and Energy Design,
Braunschweig University of Technology,
Mühlenpfordtstr. 23, 10.OG
D-38106 Braunschweig
Tel. ++49 531 - 391 3555
Fax ++49 531 - 391 8125
Germany
{fisch}@igs.bau.tu-bs.de
http://www.igs.bau.tu-bs.de

[3] Software Engineering
Department of Computer Science 3
RWTH Aachen University
Ahornstraße 55
D-52074 Aachen
{rumpe}@se-rwth.de
http://www.se-rwth.de

## Abstract

Over the last three decades comprehensive research has been carried out trying to improve commissioning processes with powerful modeling tools and methodologies for data analysis and visualization. Typically addressed application scenarios are facilities management, contracting, special consulting services and measurement & verification as part of a certification process. The results are all but convincing: Monitoring of building operation has so far not become a regular service for buildings.

We have identified a lack of process integration as a significant barrier for market success. Most methodologies have so far caused additional initial invest and transaction cost: they added new services instead of improving existing ones. The Energy Navigator, developed by synavision GmbH in cooperation with leading research institutes of the Technical University Braunschweig and the RWTH Aachen University, presents a new methodology with several new approaches. Its software platform uses state graphs and a domain specific language to describe building functions offering an alternative to the software that is so far most widely used for this task: Microsoft Word.

The Energy Navigators so called Active Functional Specification (AFS) is used for the technical specification of building services in the design phase. After construction it is completed by the supplier of the BMS (Building Management System) with the relevant sensors data as documentation of his service. Operation data can then automatically be checked for initial and continuous commissioning on whether it meets the criteria of the specification. First real life applications have shown that the Energy Navigator offers a software solution that can be applied by actually simplifying existing building processes, by improving documentation of building services and by providing an automated control of building operations.

Research shows that good operations are essential for sustainable buildings. By providing a standardized functional commissioning process and a state-of-the-art web platform for building services without significant additional cost the Energy Navigator promises to be a breakthrough for quality in building operations.

# 1. INTRODUCTION

Energy efficiency of buildings depends on their hardware – their construction, insulation, glazing, HVAC systems etc. – and on the way they are used and operated. While politics and technical guidelines as well as the historically developed procedures focused on hardware to improve energy efficiency, the software – user awareness, monitoring, commissioning – is now being moved into the spot light in Germany.

Technical guidelines like DIN EN 15232 [1] help to evaluate the potential increase in energy efficiency by using new building management technologies and algorithms. The guideline defines different BMS solutions that allow an easy A-D classification for different building services with a corresponding saving potential percentage.

Beyond the building design for the first time mandatory codes such as EnEV 2009 [2] require not only a certain level of energy efficiency on the basis of a calculated energy demand using models like defined by DIN V 18599 [3]: During the life cycle large air conditioning systems have to be inspected regularly every 10 years.

DIN EN 16001 [4] goes beyond these technical recommendations by defining an overall controlling process for building owners, public administration or businesses. It helps to establish a process of installing a monitoring system, gathering and analysing data, reporting, defining the responsibilities within the institution and the documentation of savings. The whole process is supposed to be part of a wider sustainability strategy. Its implementation is a criterion to receive tax reduction thereby providing cost savings even without immediate cuts in energy consumption. While commissioning is already strongly regulated by German building laws and defined by technical guidelines like VDI 6022 [5], DIN 15239 [6], 15240 [7] and others, it is further strengthened by certification labels by the German *DGNB – Deutsche Gesellschaft Nachhaltiges Bauen* or the *US American LEED-Standard.* Both require commissioning and monitoring actions to improve building operations including functional performance testing.

The target of any commissioning and monitoring regarding building operations is to optimize the functions of the building in operation. Although it is obvious that functions in operation must have their roots in the functional design of buildings the majority of publications in this field focuses on BMS data analysis, fault detection in operation and successful re-commissioning in the later operations. A good survey on approaches is given by Katipamula [8] and the IEA Annex 47 reports [9]. There is no research work to our knowledge that focuses explicitly on the way a functional description is actually being used after the design phase to monitor operations and serve as adaptable part of the building documentation. This aspect becomes even more important when innovative buildings are individually designed and undergo comprehensive operational adjustments in the first year of operation.

Crucial for the large scale success of monitoring buildings in operation as well as for stronger political measures requiring monitoring are cost effective solutions. Therefore the starting point for the Energy Navigator was not the technical opportunities that the internet and Building Management system (BMS) provide to analyze data. The innovative approach started by checking what is needed to ensure good building performance.

## 2. BASICS SOFTWARE TECHNOLOGIES

The Energy Navigator is a technical state of the art software platform, using client- server mechanisms, providing the possibility to not only monitor but also specify the behavior of a BMS. The platform provides all means necessary to get a closed circuit between specification, monitoring, data analyses and optimization. Monitoring of buildings and providing sophisticated analyses means handling mass data. The platform is able to collect either automatic, by using OPC to directly read values from the BMS, or manual, by importing data from a data logger, in discrete intervals. Since we created a highly adaptable platform we are able to scale up the data collection to one data point each second for a single sensor. But under normal conditions a resolution of collecting one data point every 15 minutes for a single sensor has proven to be feasible and state of the art in building automation systems. Even this coarse grained data collection leads to a lot of information that needs to be stored and processed by the platform. Consider a building having 1000 sensors each producing a data point every 15 minutes. Thus we get 96000 data points a day from a single building being about 35 million data points a year. To keep the performance and scalability we use cloud computing based storage techniques. Additionally most modern building automation systems provide the possibility to not only log data points but to log also markers signaling different modes a building or part of it resides in.

Each data collection is followed by a preprocessing step ensuring that all the values have equidistant timestamps and filtering outlier. Having collected all the data the user needs to be able to create custom analyses for the monitored building. To support these tasks the Energy Navigator platform provides the possibility to create several elements aiding the analyses of mass data measured in a building. We therefore created a Domain Specific Language (Karsai et al., 2009 [13]) for specifying and modeling buildings. To create such a language we use our framework MontiCore (Krahn et al., 2010 [14]). The language and basic application principles have been described in (Fisch, Plesser et al., 2010, 2011 [15]).

The Energy-Navigator Platform is divided into a client side that can be web based or desktop based. The web based client concentrates on visualization of reports for end users of a building (management reports, public awareness). The rich client is used by energy experts that specify, evaluate and validate the proper operation of a building. The platform is designed for multiuser access, figure 1.

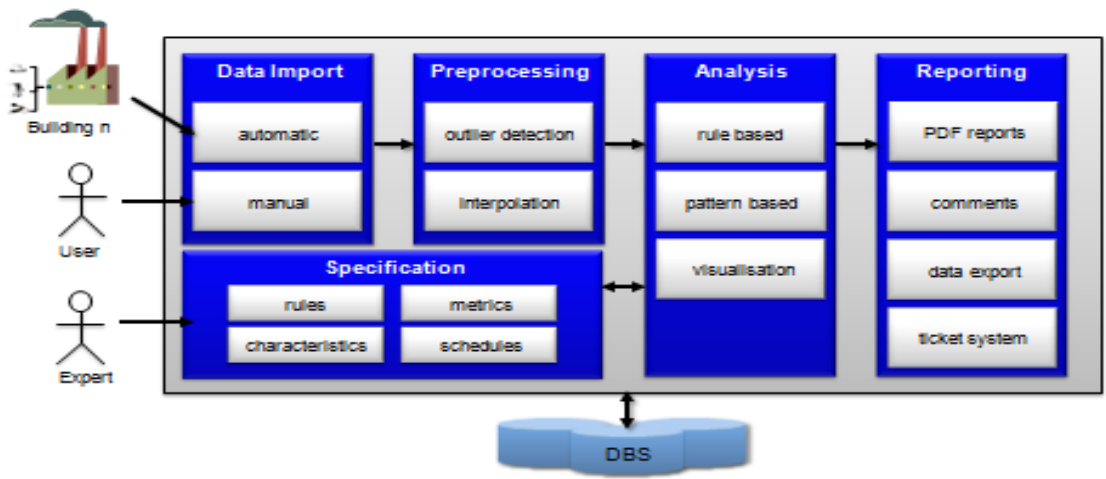

Figure 1 Software architecture of Energy Navigator platform

On the server side there are several components forming the backend platform. Each component can be plugged in, so that the platform is highly adoptable to build complex solution for each building. The components are grouped into Data Import, Preprocessing, Analysis, Report and Specification. The platform offers automatic or manual (file based) import depending on the data availability inside the building system. After the import data can be automatically preprocessed. This is a crucial step to guarantee a high data quality for analysis and further calculations. There is outliner detection or interpolation of missing values, a transformation to equidistant time steps, to mention only some preprocessing workflows. The analysis can be done automatically by using a formal specification model and intelligent algorithms. Additionally an energy expert can use the platform to visualize the imported data with multiple plot types, e.g. line plots, scatter plots or carpet plots. Another feature of the Energy-Navigator Platform is the reporting component. With this component we are able to automatically create reports and inform other users about the current system status, and potential performance issues of the building.

The platform uses a database backend and high efficient algorithms that are optimized to handle mass data of current building systems. The data can be processed and stored localized, e.g. for reasons of data privacy protection, or with innovative cloud computing technologies for better scalability and resource sharing.

# 3. ACTIVE FUNCTIONAL DESCRIPTIONS USING THE DOMAIN SPECIFIC LANGUAGE

A crucial aspect is how to deal with complexity in the design of buildings. The Energy Navigator platform establishes the concept of templates for every artifact. The idea behind this concept is that an expert can specify his/her knowledge once at the beginning and use these templates easily for every building that he/she operates or manages. For a concrete building the expert adds the templates to the workspace and maps the concrete sensors to the template. The use of a library is a key feature for reusability of expert knowledge.

The main concept of the Energy Navigator platform is a constraint language that is used to specify facilities, systems and building operation. The constraint language is an adapted variant of the Object Constraint Language (OCL), part of the well-known Unified Modeling Language (UML). [16] The language is developed with MontiCore [17] -a framework for efficient language design. The main artifacts of the language are rules, functions, metrics, time routines and characteristics. The concepts can be described as follows:

Rules

Rules are logical and arithmetical expressions [18] that can be evaluated to Boolean values true or false. They are defined in the context of sensors and can be used to specify the desired behavior of a system.

A rule may contain logical operators like AND, OR, IMPLIES, NOT, IF-THEN-ELSE etc. and arithmetical operators like PLUS, MINUS, MULTIPLY, DIVIDE etc. An important concept to deal with complexity of such specifications is referencing sub rules or other language elements, like functions from a library (e.g., MAXIMUM, SUM, AVERAGE etc.). The referenced elements are specified self-contained in separate artifacts to enable reusability within other language artifacts.

After specifying a rule an automated analysis can be executed by the platform. For each rule (and sub element) a virtual sensor is created. A virtual sensor is analogous to real sensors an equidistant sequence of values. In the case of a rule a valid value can be true, false, missing or undefined. A result is missing if no context information (sensor data) is given. An undefined result means that no evaluation was possible. The resulting sequence has the same temporal resolution as the context sensors, for instance 15 minutes.

Functions

Functions are very similar to rules but they are not resulting in Boolean but numeric values. The context of a function is one or more real or virtual sensor. An example with two sensors is the function $f(s1, s2) = s1 - s2$ where s1 could be the supply temperature and s2 the return temperature in a heating circuit. The function calculates the temperature spread, which can directly be used for visualization or can be referenced in other functions, rules or metrics.

### Characteristics

Characteristics are a powerful concept to specify relations between two sensor dimensions. A single characteristic or an upper and lower characteristic can be specified by a point based graphical editor. The characteristic can be defined as a function or as a rule.

### Metrics

Functions, rules and characteristics are equidistantly processes, what means that the value sequences are iterated from timestamp to timestamp and results are calculated by the given parameters. Compared to functions metrics are calculating values for a given time span for example the energy consumption of one month or an average daily temperature.

A so called base metric consists of a context sensor, for example a temperature measurement every 15 minutes. Additionally a base function can be added, for example Average, Minimum, Miximum, Sum etc. Last but not least a time quantization can be added. At this stage the system supports Day, Week, Month, Quarter and Year.

With this simple configuration mechanism a lot of standard metrics can be calculated. Additionally customized metrics can be used from a library which can be parameterized, e.g. the calculation of a standard deviation. Metrics are very helpful for management reports.

### Time Routines

Time routines can be used to differentiate between several operations modes. An example for use is the specification of a public school building, where facilities should have different operation modes for weekdays, weekends or holidays. A time routine can be defined a set of time ranges. Values for year, month, day, hour, minute, second and weekday can summarized to a schedule. It is also possible to include or exclude additional time routines to specify exception, e.g., holidays.

### Ticket System

The ticket system is highly adoptable for each building. It can be used for our automatic and manual analysis of the building data. After a rule was evaluated by our system and failed because the facility is not running conform to the specification, we create a new ticket that holds all necessary information for the facility manager. So violated constrains or specifications from the design phase can be traced during the operation of a building.

### Visualization

Our platform offers several types of visualization. We can use line plots, scatter plots and carpet plots as shown in figure 2, figure3 and figure 5.

### Reporting

After the analysis we offer the automated generation of reports. These reports are customizable. It is possible to insert handwritten parts to the generated reports. So an energy expert can create a special purpose report which focuses only on the heating facilities of a building that consists only of the important sensors, facilities, rules and visualized plots. This report can then be reused for a specific building. The expert can make additional comments, e.g. to give suggestions what should be changed for the operation of the heating facilities. These comments are saved independent of the generated report. If the report is regenerated, e.g. after one month new operational data of the building is available; the original handwritten comments are still there.

## 4. NEW WAYS TO DESIGN AND REPORT FUNCTIONAL QUALITY

Conventional analysis and reporting rely on operation data only. Typical graphs of energy information simply state the energy consumption of buildings for past periods of time, sometimes aggregated to create a kind of energy accounting, figure 2.

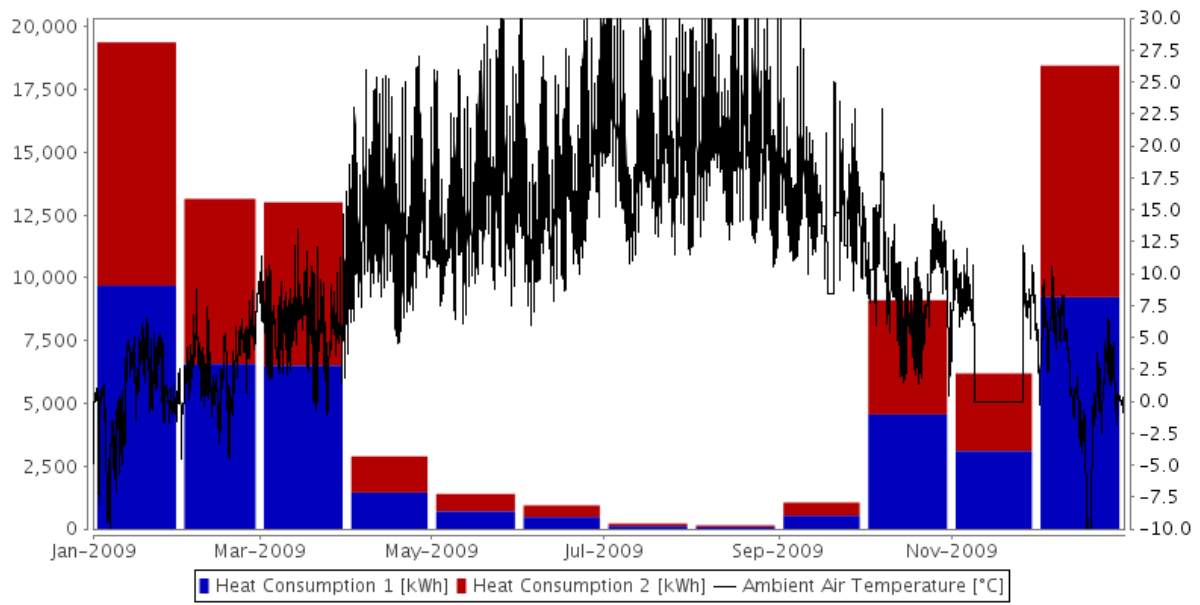

Figure 2              Example graph: historic data for monthly energy consumption

More sophisticated tools for data analysis provide graphs to visually examine large amounts of data in short time. The power of these tools in practical application, as shown in figure 3, depend heavily on the expert knowledge and motivation of the user since they need precise technical understanding and interpretation.

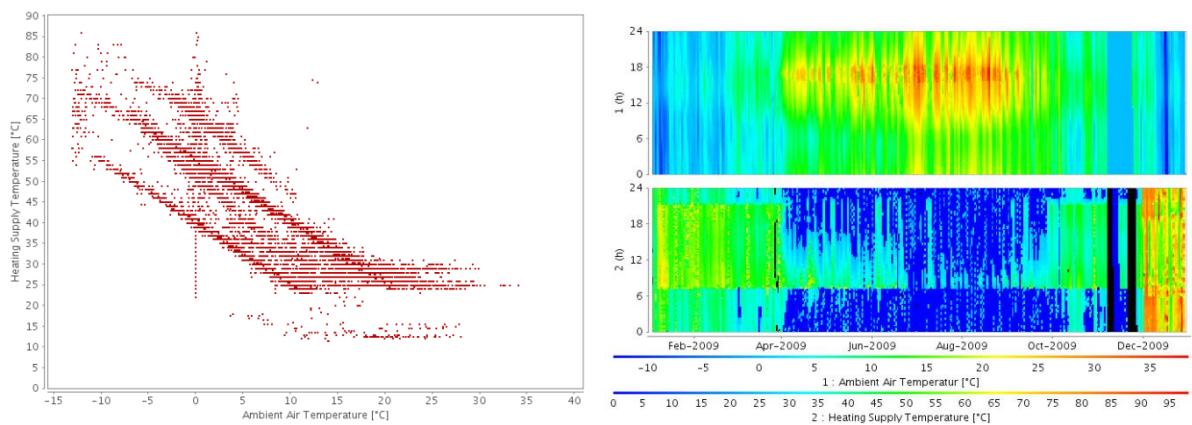

Figure 3              Example graph: visualization techniques for easy analysis of building operations

Advanced tools use modeling techniques to define a target or limit value. Most common are *black box models*, usually based on historic data from the building or system that has been transformed into predicted values for the operation data by statistic modeling, or *white box models* in which physical and technical knowledge is applied in dynamic simulation tools to model a target value. Most common are target values for the total heat and electricity consumption of buildings. Easy models can also be used for simple limit checking e.g. of load curves or temperatures.

The Energy Navigator uses the active functional specification (AFS) as modeling methodology using the domain specific language in combination with technical schematics of the corresponding system as shown in figure 4.

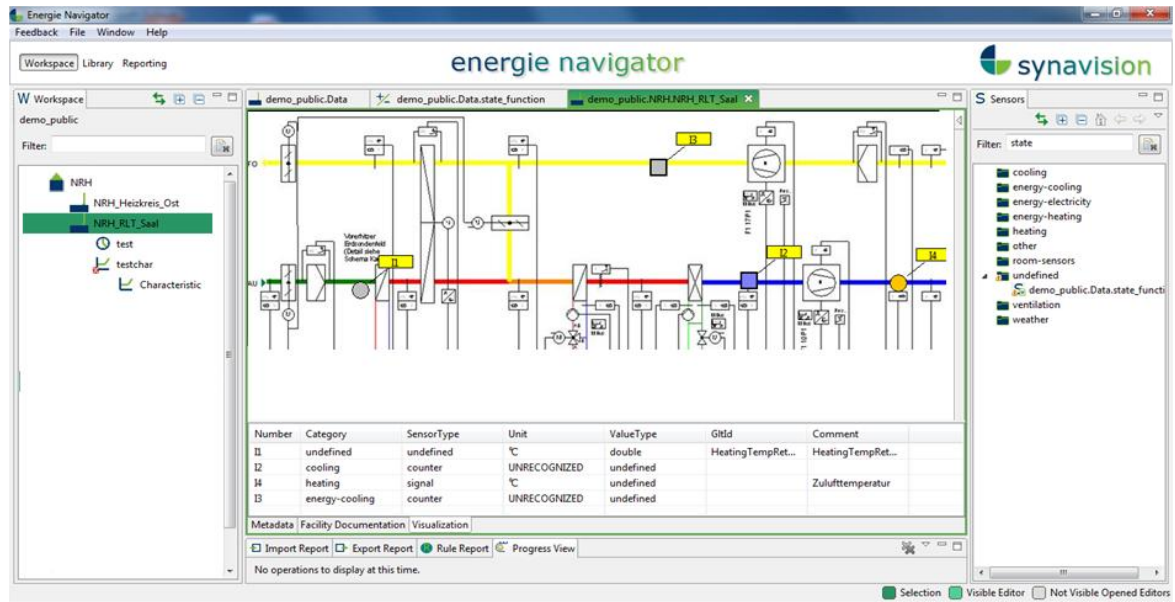

Figure 4　　　　　　　Active Specification of the Energy Navigator platform

The graphical editor helps to easily match the design specification and the operation data. The concept has some promising advantages:

- **The specification is not only used for the "checking" in operation but already for the "explaining" of the intended** function in the design phase. This improves communication within the project especially in complex and individual buildings.

- **Since the specification is "online" it can be expanded and adjusted during the life cycle** keeping the documentation of the building up to date.

- By defining a target value in the design phase the building owner can not only check operation later on automatically. The specified function can be part of the contract for the BMS contractor: performance becomes a measureable part of the service.

Although all the graph that have been mentioned above are also possible, the key graph shows red or green for each point of time indicating correct or incorrect performance compared to the specification, figure 5.

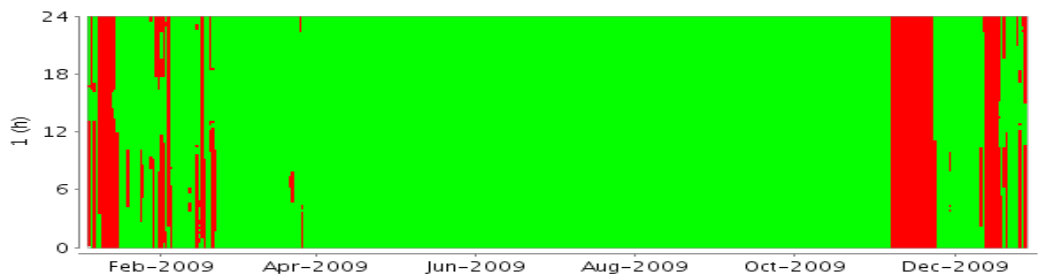

Figure 5　　　　　　　Checking the performance: green indicates correct performance, red indicates malfunction

In addition to the graph the information behind it can be used to create tickets, emails, SMS or any other kind of reporting with web 2.0 technologies.

## 5. TURNING BUILDING CONCEPTS INTO BUILDING PERFORMANCE

The Energie-Navigator provides new opportunities for buildings in design, construction and operation. The platform is web-based, easy to apply and its application therefore feasible for all partners during the whole life cycle. Since all partners work in the same environment, the system guarantees means for a continuous use thereby bridging the gap between design and operation phase, figure 6.

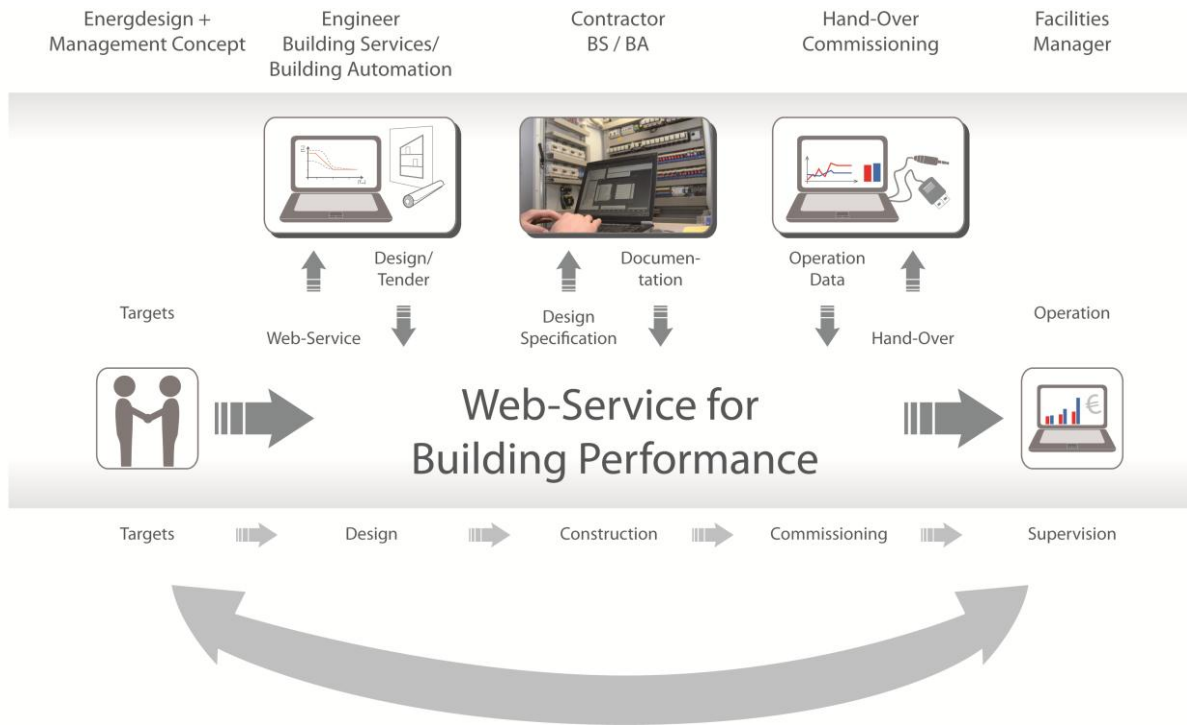

Figure 6   Project workflow for the Energy Navigator

By integrating a powerful tool into the existing workflows in today projects and providing a language for the specification of building functions the Energy Navigator promises to powerfully help to turn energy efficient building concepts into truly energy efficient buildings.

We are finishing the initial testing of the application and have founded the synavision GmbH (www.synavision.de) that will bring the Energy Navigator on the market as a software product in 2012.